\documentclass[twocolumn,showpacs,nofootinbib,preprintnumbers,prd]{revtex4-1}

\usepackage{amsmath}
\usepackage{amsfonts}
\usepackage{amssymb}
\usepackage{graphicx}
\usepackage[titletoc]{appendix}
\usepackage{color}
\usepackage{hyperref}
\usepackage{cleveref}
\usepackage[rightcaption]{sidecap}
\usepackage{subcaption}
\usepackage{comment}

\usepackage{mathtools}

\usepackage{dcolumn}

\usepackage{array}
\usepackage{ctable}
\usepackage{multirow}
\usepackage{siunitx}

\usepackage{tabularx}
\usepackage{natbib}
\usepackage{booktabs}
\usepackage{supertabular}
\usepackage{verbatim}

\usepackage{cancel}
\usepackage{graphicx}
\usepackage{dcolumn}
\usepackage{bm}
\usepackage{braket}

\def\prd{Physical Review D}
\def\mnras{Monthly Notices of the Royal Astronomical Society}

\def\apj{The astrophysical journal}

\def\aap{Astronomy \& Astrophysics}

\begin{document}

\title{Exploring the cosmic microwave background dipole direction using gamma-ray bursts}

\author{Orlando Luongo}
\email{orlando.luongo@unicam.it}
\affiliation{Universit\`a di Camerino, Via Madonna delle Carceri, Camerino, 62032, Italy.}
\affiliation{Department of Nanoscale Science and Engineering, University at Albany-SUNY, Albany, New York 12222, USA.}
\affiliation{Istituto Nazionale di Astrofisica (INAF), Osservatorio Astronomico di Brera, 20121 Milano, Italy.}
\affiliation{Istituto Nazionale di Fisica Nucleare (INFN), Sezione di Perugia, Perugia, 06123, Italy,}
\affiliation{Al-Farabi Kazakh National University, Al-Farabi av. 71, 050040 Almaty, Kazakhstan.}

\author{Marco Muccino}
\email{marco.muccino@unicam.it}
\affiliation{Universit\`a di Camerino, Via Madonna delle Carceri, Camerino, 62032, Italy.}
\affiliation{Al-Farabi Kazakh National University, Al-Farabi av. 71, 050040 Almaty, Kazakhstan.}
\affiliation{ICRANet, Piazza della Repubblica 10, 65122 Pescara, Italy.}

\author{Francesco Sorrenti}
\email{francesco.sorrenti@unige.ch}
\affiliation{D\'epartement de Physique Th\'eorique and Center for Astroparticle Physics,\\
Universit\'e de Gen\`eve, 24 quai Ernest  Ansermet, 1211 Gen\`eve 4, Switzerland.}

\date{\today}

\begin{abstract}
We search for dipole variations in the Hubble constant $H_0$ using gamma-ray burst (GRB) data, as such anisotropies may shed light on the Hubble tension. We employ the most recent and reliable GRB catalogs from the $E_{p}–E_{iso}$ and the $L_0-E_{p}-T$ correlations. Despite their large uncertainties, GRBs are particularly suited for this analysis due to their redshift coverage up to $z\sim9$, their isotropic sky distribution that minimizes directional bias, and their strong correlations whose normalizations act as proxies for $H_0$. 
To this aim, a whole sky scan -- partitioning GRB data into hemispheres --
enabled to define dipole directions by fitting the relevant GRB correlation and cosmological parameters. 
The statistical significance across the full $H_0$ dipole maps, one per correlation, is then evaluated through the normalization differences between hemispheres and compared against the CMB dipole direction. 
The method is then validated by simulating directional anisotropies via Markov Chain Monte Carlo analyses for both correlations.
Comparison with previous literature confirms the robustness of the method, while no significant dipole evidence is detected, consistently with the expected isotropy of GRBs. This null result is discussed in light of future analyses involving larger datasets.
\end{abstract}

\maketitle
\tableofcontents

\section{Introduction}\label{sec:1}

The concordance $\Lambda$CDM background is the current most suitable cosmological paradigm able to provide satisfactory fits to both early- and late- time observations \cite{Peebles:2024txt,Blanchard:2022xkk}. Even though successful, the concordance paradigm is jeopardized by yet unsolved issues \cite{Bull:2015stt}, such as the recent findings by the DESI Collaboration \cite{DESI:2025zgx}, indicating evolving dark energy\footnote{The standard $\Lambda$CDM model appears still a viable approximation to describe the background, since alternative findings show that DESI data do not show a huge discrepancy with respect to our current understanding \cite{Luongo:2024fww,Carloni:2024zpl}, as sometimes claimed \cite{Wang:2024rjd}.
} and, even more accepted, the so-called cosmic tensions \cite{2022JHEAp..34...49A, 2021APh...13102605D, 2021APh...13102604D,2020PhRvD.102b3518V, 2021MNRAS.505.5427N, 2023Univ....9..393V}.

A well-known open challenge is the so-called Hubble tension, currently at the $4.1\sigma$ level, between the Hubble constant by the SH0ES Collaboration, $H_0 = (73.04 \pm 1.04)\,\mathrm{km\,s^{-1}\,Mpc^{-1}}$, based on local Cepheid-calibrated type Ia supernovae (SNe Ia) \cite{2022ApJ...934L...7R}, and the one by the Planck Collaboration, $H_0 = (67.36 \pm 0.54)\,\mathrm{km\,s^{-1}\,Mpc^{-1}}$, under the assumption of the $\Lambda$CDM model using cosmic microwave background (CMB) data \cite{Planck:2018nkj}.

A further tension, still arising by comparing local universe observations with CMB measurements, is dubbed \textit{cosmic dipole} \cite{Secrest:2025wyu,Yoo:2025qdq}.
The cosmic dipole refers to as a dipole in temperature,  $\Delta T/T \sim 10 ^{-3}$, showing a hot area likely placed in the northern galactic hemisphere. The dipole itself may correspond to the \emph{peculiar motion of the observer with respect to the CMB frame} and for this reason is also known as \emph{kinetic dipole} \cite{Kogut:1993ag,Planck:2013kqc,Planck:2018nkj,Saha:2021bay}. The dipole amplitude $A_\star=(1.23000\pm0.00036)\times10^{-3}$ is measured in the direction, labeled with a star, $(l_\star,b_\star)=(264.02^\circ\pm0.01^\circ,48.253^\circ\pm0.005^\circ)$ in galactic longitude and latitude, respectively \cite{Planck:2018nkj}, or 
\begin{equation}
(\alpha_\star,\delta_\star) =(167.942\pm 0.007, -6.944\pm 0.007),
\label{e:d-Planck-solar}
\end{equation}
in the equatorial coordinates, with right ascension (RA) $\alpha$ and declination (DEC) $\delta$. 

The cosmological principle disagrees with local results as  isotropy is expected to emerge only on sufficiently large scales, typically beyond $\sim 100$ Mpc. There, the matter distribution, traced by galaxies and galaxy clusters \cite{plionis,Rowan_Robinson_2000,Plionis:1997wj, scaramella,Kocevski:2004pf}, appears statistically homogeneous. 
Hence, at such scales, one would expect the cosmic dipole to converge toward the CMB kinematic dipole \cite{Ellis:1984}. 

Quite unexpectedly, this is not completely the case, since dipole measurements from diverse cosmological probes, including SNe Ia, quasars, radio sources, and gamma-ray bursts (GRBs), consistently reveal enhanced dipole amplitudes -- and at times even mild directional discrepancies -- relative to the CMB dipole. 
In a recent work \cite{2022PhRvD.105j3510L}, this mismatch showed an increase of $H_0$ along a certain direction, indicating the evidence for the cosmological principle breakdown, albeit depending on the kind of data points employed into the computation. 
Accordingly, one can wonder whether the cited mismatch points to a possible intrinsic anisotropic component beyond the purely kinematic contribution, see e.g.~Refs. ~\cite{Carr_redshift_pantheon+,2020A&A...643A..93H,Rubart:2013tx,Colin:2017juj,Tiwari:2016,Colin:2019opb,Siewert:2021,singal,Secrest:2020has,Bengaly:2017slg,Siewert:2020krp,2025A&A...694A..77L}.

Thus, if confirmed, the intrinsic dipole would imply the \emph{violation of the cosmology principle of isotropy and homogeneity}, opening novel avenues towards new physics \cite{Wagenveld:2025ewl,Chen:2025rru,Panwar:2025kkd}.
The origin of this  intrinsic anisotropy as well as its properties are not fully clear and, as a possible theoretical explanation, it has been proposed that the anisotropy may be due to the back-reaction of cosmological perturbations, mimicking dark energy effects \cite{Kolb:2005da,Bolejko:2016qku}.  

In this paper, we focus on the $H_0$ dipole, namely on searching for variations of the Hubble constant across the sky and along particular directions, following the procedure of Ref.~\cite{2022PhRvD.105j3510L}.  Indeed, the possible existence of a dipole anisotropy in $H_0$ is a fascinating topic, that may even contribute to resolve the well-known Hubble tension.
To this aim, we here utilize the most updated and accurate GRB catalogs from two different correlations: the first is the well-consolidated $E_{p}-E_{iso}$ or Amati correlation \cite{2002A&A...390...81A,AmatiDellaValle2013,2021JCAP...09..042K}, while the second is known as the $L_0-E_{\rm p}-T$ or Combo correlation \cite{2015A&A...582A.115I,Muccino:2020gqt}, a widely-used as a robust alternative to the first one that consists of a combination of prompt and X-ray afterglow observables.

Albeit affected by larger uncertainties with respect to other cosmic probes, GRB catalogs are the most suited for dipole searches in view of the reasons listed below.
\begin{itemize}
\item[-] Both catalogs cover redshift up to $z\sim 9$, which are ideal to test the validity of the cosmic principle\footnote{For completeness, GRBs cannot provide insight in peculiar motions in the local Universe in view of the lack of low-$z$ sources.}.
\item[-] Unlike SNe Ia and quasars, GRB distribution is isotropic \cite{2022PhRvD.105j3510L}. This, in principle, may cancel possible directional bias, where the sources of the catalog mostly cluster. Moreover, their distribution lies on intermediate redshift domains and, thus, may in principle depart from the background dynamics. 
\item[-] In both the $E_{p}-E_{iso}$ and he $L_0-E_{\rm p}-T$ correlations, the  intercepts $a$ degenerate and positively correlate with $\log H_0$, thus, representing proxies for $\log H_0$. In particular, any deviations $\Delta a$ would correspond to deviations $\Delta H_0/H_0$ along a particular direction of the sky \cite{2022PhRvD.105j3510L}.
\end{itemize}

We perform two Markov chain Monte Carlo (MCMC) analyses, one per correlation, where -- to explore variations of $H_0$ across the sky -- we scan over a grid of $j$ angular directions $({\rm RA},{\rm DEC})$, each covering a $12^\circ\times12^\circ$ patch, and define for each of them a normal vector that splits each catalog in hemispheres. Correspondingly, ensuring that each hemisphere shall fulfill the same
adopted correlation, we determine along each direction the correlation and the cosmological parameters. Afterwards, focusing on the variation between hemispheres of the correlation normalization (as a proxy of $H_0$), we compute the projection map of the corresponding significance for such variations over all the points of the grid and compare it with the CMB dipole direction. 
Further, we test our procedure with mock catalogs with artificial dipoles injected within and with previous results obtained utilizing similar GRB data \cite{2022PhRvD.105j3510L}. 
Our results do not evidence a net dipole. 
Rather, as expected from the GRB isotropy, we find no net evidence for the dipole itself. This result is therefore commented throughout the text in view of future efforts, making use of further data points. 

This paper is organized as follows. In Sec.~\ref{dip_data}, we review some determinations of the dipole amplitudes and directions obtained from SNe Ia, quasars, radio galaxies, and GRBs.
In Sec.~\ref{sec:GRB_dip} we introduce our method for assessing the existence of a cosmic dipole, in the form of a $H_0$ dipole, in both selected GRB catalogs and show the corresponding results.
In Sec.~\ref{concl} we draw our physical outcomes, emphasizing the main results of our method and explaining the consequences of our analysis. Moreover, we draw our conclusions and perspectives.

\section{Dipole determinations across cosmic data sets}
\label{dip_data}

Several recent studies have investigated the presence of dipole anisotropies (and also additional terms) to test the cosmological principle, which posits that the universe is homogeneous and isotropic on large scales. 

\paragraph*{{\bf SNe Ia.}} Dipoles in the luminosity distance of SNe Ia \cite{Sorrenti:2022zat} were searched in the \textit{Pantheon}+ sample -- a dataset of $1701$ objects, $77$ of which located in a galaxy hosting a Cepheid \cite{pantheon_cosmo}. 
The results showed the presence of a dipole with amplitude broadly consistent with that of the CMB dipole, albeit with a discrepant direction (at $\sim 3\sigma$ level). This discrepancy was accounted for by assuming a local bulk motion that aligns with the \textit{a priori} correction applied by the Pantheon+ collaboration in their cosmological analysis~\cite{Carr_redshift_pantheon+} and it is consistent with the results found in the CosmicFlows4 dataset \cite{watkins23, Whitford:2023oww}.
Further investigations on peculiar velocities, led to the discovery of significant quadrupole and, at very low redshift, negative monopole terms, with amplitudes comparable to that of the dipole associated with the bulk motion \cite{Sorrenti:2024a}.

These results underline the importance of local peculiar velocities in determining the correct cosmic dipole using SNe Ia.

\paragraph*{{\bf Quasars.}} Using hemisphere comparison and dipole fitting methods on combined Pantheon and quasar data sets, a maximum anisotropy level of $A = 0.142 \pm 0.026$ in the direction $(l, b) = (316.08^\circ, 4.53^\circ)$ was found. However, the statistical significance was modest, around $1.23\sigma$, suggesting that the observed anisotropy could be due to statistical fluctuations or systematic effects \cite{2020A&A...643A..93H}.
Similarly, in Ref.~\cite{2024JCAP...11..067A}, it was concluded that the quasar dipole is consistent with the CMB dipole, indicating no significant deviation from isotropy.

On the other hand, other works evidenced tensions with the cosmic dipole. 
For example, a Bayesian analysis performed on the quasar distribution of the Quaia sample led to the discovery of a dipole inconsistent with the CMB dipole \cite{2024MNRAS.527.8497M}.
A further evidence for an intrinsic dipole was inferred from quasar counting in the CatWISE2020 catalogue, yielding an amplitude of $1.5\times10^{-2}$ in tension with the CMB dipole at $4.9\sigma$ level \cite{Secrest:2020has}.

Finally, an independent evidence for a dipole in quasars was obtained in Ref.~\cite{2022PhRvD.105j3510L}, where it was claimed that $H_0$ is larger in the CMB dipole direction or along aligned directions.
This result got by using $H_0$, which \textit{a priori} has no directional preference, is thus independent from any anisotropy and poses hints that the dipole anisotropy might be related to the ongoing Hubble tension \cite{2022ApJ...934L...7R,Planck:2018nkj}.

\paragraph*{{\bf Radio Galaxies.}} Investigations into the distribution of radio galaxies have detected a cosmic radio dipole, an anisotropy in the number counts of radio sources. This dipole is generally consistent in direction with the CMB dipole, but the inferred velocity is much larger. 
Moreover, measurements from number counting of radio galaxy catalogues, such as the TIFR GMRT Sky Survey (TGSS), the NRAO VLA Sky Survey (NVSS), and the Westerbork Northern Sky Survey (WENSS), indicate that the radio dipole amplitude is in the range $0.010$--$0.070$, several times larger than the CMB dipole, with an uncertainty in the order of $10^{-3}$ 
\cite{Blake:2002gx,2011ApJ...742L..23S,Rubart:2013tx,Gibelyou:2012ri,Tiwari:2013vff,Fernandez-Cobos:2013fda,Tiwari:2013ima,Tiwari:2016,Bengaly:2017slg,Colin:2017juj,Colin:2019opb,Siewert:2020krp,Siewert:2021}.

These results raise questions about the underlying causes and suggest the need for further investigation into potential systematic effects or new physics \cite{2023A&A...675A..72W}.

\paragraph*{{\bf GRBs.}} The angular distribution of GRB data, particularly from the FERMI/GBM catalog, has been studied for testing isotropy. While the overall distribution of GRBs in the sky appears isotropic, analyses of their fluence have revealed a dipolar pattern that suggests that there may be anisotropies in their energy output, potentially pointing to underlying astrophysical processes \cite{2025A&A...694A..77L}.

Like in the case of quasars, $H_0$ was also used in GRBs to test the presence of a dipole~\cite{grb1, grb2}, that resulted for selected burst subsamples, specially for those directed around the CMB dipole direction \cite{2022PhRvD.105j3510L}.

Finally, GRBs were used to test the dipole anisotropy in the dipole-modulated $\Lambda$CDM and Finslerian cosmological models \cite{2022MNRAS.511.5661Z}.
The weak deduced anisotropy has a direction consistent with that got from the Pantheon SN Ia catalog, but with smaller uncertainties.  
Moreover, when combined with SNe Ia, it was shown that GRBs considerably impacted the results from the Pantheon sample.

\section{Search for a $H_0$ dipole in GRB data}
\label{sec:GRB_dip}

Following the methodology illustrated in Ref.~\cite{2022PhRvD.105j3510L}, we utilize two catalogs of GRBs:
\begin{itemize}
    \item[-] the A118 catalog consisting of $\mathcal N_A=118$ GRBs with redshift range $z\in[0.3399,8.2]$, fulfilling the well-established $E_{p}-E_{iso}$ or Amati correlation \cite{2002A&A...390...81A,AmatiDellaValle2013} with a small intrinsic dispersion \cite{2021JCAP...09..042K};
    \item[-] the C182 catalog composed of $\mathcal N_C=182$ bursts in the redshift range $z\in[0.0368,9.4]$, fulfilling the $L_0$--$E_{\rm p}$--$T$ or Combo correlation \cite{2015A&A...582A.115I,Muccino:2020gqt}. 
\end{itemize}

It is worth recalling that, differently from quasars which exhibit clear clustering in specific regions of the sky (in particular around the CMB direction), GRBs are isotropically distributed albeit with catalogs smaller -- so statistically less powerful -- than quasar sample.

\subsection{The $E_{p}-E_{iso}$ correlation data set}

The $E_{p}-E_{iso}$ correlation \cite{2002A&A...390...81A,AmatiDellaValle2013} relates the GRB intrinsic spectral peak energy $E_{\rm p}$ (in keV units) and the isotropic energy $E_{\rm iso}$ (in erg units)
\begin{equation}
\label{eq1}
\log E_{\textrm{iso}} = a + b \log E_{\textrm{p}},  
\end{equation}
where the free parameters $a$ and $b$ are the normalization and the slope, respectively. The correlation is also characterized by an intrinsic dispersion parameter $\sigma_{\rm ex}$. Ideally, if $a$ and $b$ are known \emph{a priori}, the isotropic energy, $E_{\rm iso}$ could be directly evaluated from Eq.~\eqref{eq1}. In reality, it is a cosmology-dependent quantity, $\mathcal E_{\rm iso}$, that has to be computed from the measured bolometric fluence $S_{\rm b}$ via
\begin{equation}
\label{eq2}
\log \mathcal E_{\rm iso} = \log(4 \pi)+ 2\log D_{L}(z) + \log S_{\textrm{b}} -\log (1+z),
\end{equation}
where, within the flat $\Lambda$CDM model, the luminosity distance, $D_{\rm L}(z)$, can be expressed as 
\begin{equation}
\label{Dlambda}
D_{L} (z) = \frac{c}{H_0} (1+z) \int_{0}^{z} \frac{{\rm d} z'}{\sqrt{1-\Omega_m + \Omega_m(1+z')^3}}, 
\end{equation}
and depends only upon the matter density parameter $\Omega_m$ as we fix $H_0 = 70$ km/s/Mpc. 
It is clear from Eqs.~\eqref{eq1}--\eqref{Dlambda} that $a$ degenerates and positively correlates with $\log H_0$, thus, $a$ represents a proxy for $\log H_0$.
Moreover, fixing $H_0$ does not introduce any bias in our analysis because, any deviations $\Delta a$ correspond to deviations $\Delta H_0/H_0$ along a particular direction of the sky.

\begin{figure*}[t]
\centering
\includegraphics[width=0.75\hsize,clip]{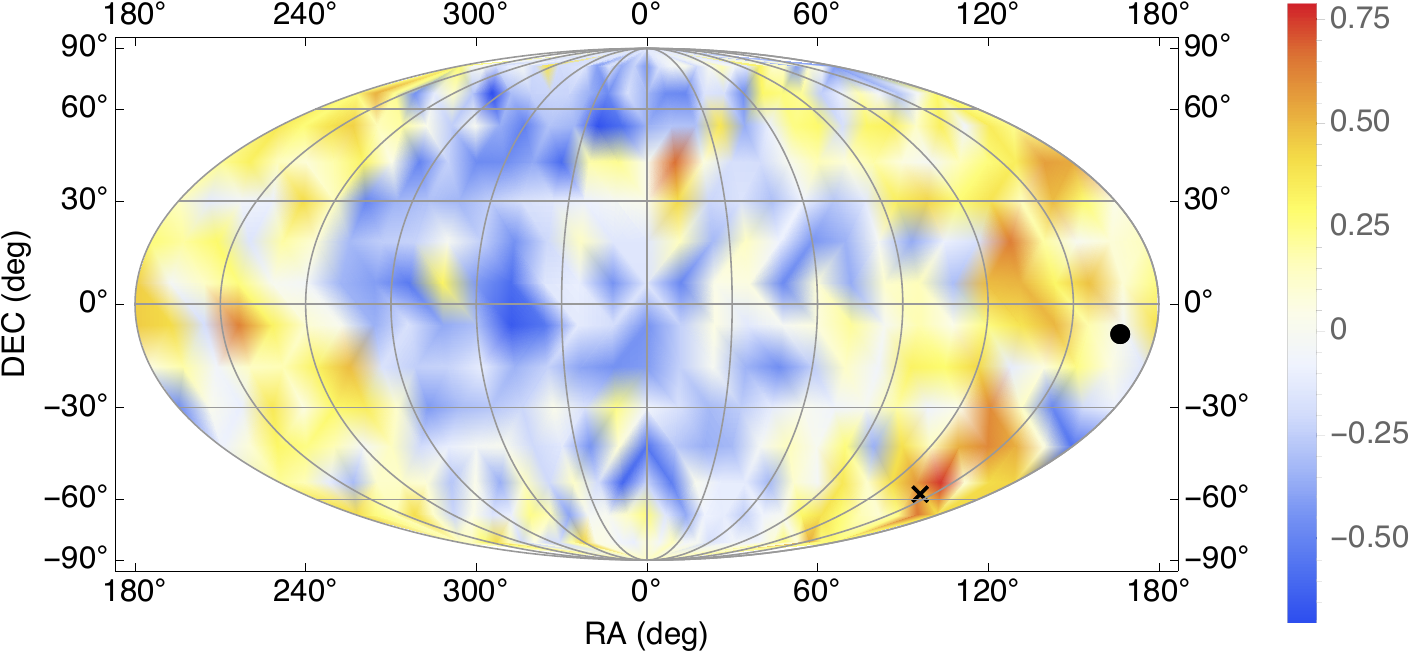}
\includegraphics[width=0.75\hsize,clip]{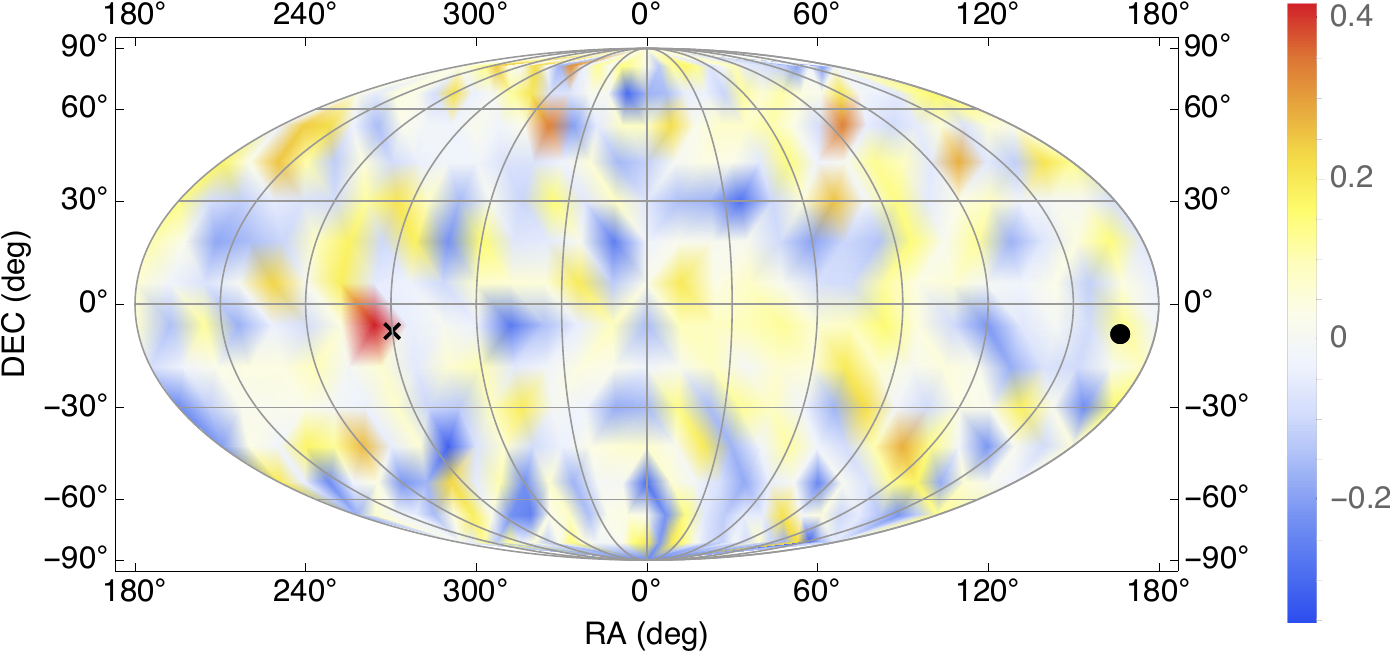}
\caption{Mollweide RA--DEC projection maps of the significance $\sigma$ (see right side color-coded bars) computed from Eq.~\eqref{sigma} for the A118 (top) and the C182 (bottom) catalogs. The maximum $\Delta a$ direction $(\alpha_{\rm m},\delta_{\rm m})=(144^\circ,-54^\circ)$ for the A118 catalog (top panel, black cross), the maximum direction  $(\alpha_{\rm m},\delta_{\rm m})=(270^\circ,-6^\circ)$ for the C182 catalog (bottom panel, black cross), and the CMB dipole direction $(\alpha_\star,\delta_\star) = (168^{\circ}, -7^{\circ})$ (black dots) are also shown.}
\label{fig1}
\end{figure*}

For uncalibrated correlations, the best-fit parameters include also $\Omega_m$ that can be determined, together with $a$, $b$ and $\sigma_{\rm ex}$, by maximizing the log-likelihood function  
\begin{align}
\label{L2}
\mathcal F_A = - \frac{1}{2} \sum_{i=1}^{\mathcal N_A} \! \left[\! \frac{ \left(\log \mathcal E_{\rm iso,i}\!-\!\log E_{\textrm{iso}, i}\right)^2}{\sigma_A^2}\! + \ln(2 \pi \sigma_A^2 ) \!\right]\!, 
\end{align}
where $\sigma_A^2 = \sigma_{\log S_{\rm b,i}}^2 +  b^2 \sigma_{\log E_{\rm p,i}}^2 + \sigma_{\rm ex}^2$ is the global error of the correlation, and $\sigma_{\log S_{\rm b}}$ and $\sigma_{\log E_{\rm p}}$ are the errors on the logarithms of $S_{\rm b}$ and $E_{|rm p}$, respectively.
The MCMC fit performed on the A118 data set provides the best-fit results summarized in Table~\ref{tab1}.

\subsection{The $L_0$--$E_{\rm p}$--$T$ correlation data set}

The $L_0-E_{\rm p}-T$ correlation considers both quantities from the prompt and X-ray afterglow emissions \citep{2015A&A...582A.115I,Muccino:2020gqt}. It is obtained by combining the Amati correlation with the $E^X_{iso}-E_{iso}-E_p$ correlation \citep{2012MNRAS.425.1199B}, leading to
\begin{equation}
\label{l00}
\log L_0 = a +b\log E_{\rm p} - \log T,
\end{equation}
with normalization $a$, slope $b$, and intrinsic dispersion $\sigma_{\rm ex}$. 
The correlation links $E_{\rm p}$ from prompt emission (in keV units) with X-ray afterglow effective rest-frame duration of the plateau $T$ (in s units) and the plateau luminosity $L_0$ (in erg/s units). 
Again, if $a$ and $b$ are known \emph{a priori}, the luminosity $L_0$ could be directly calculated from Eq.~\eqref{l00}. In reality, it has to be evaluated from the measured flux $F_0$ (in erg/cm$^2$/s units) through the luminosity distance from Eq.~\eqref{Dlambda} via
\begin{equation}
\label{l0}
\log \mathcal L_0 = \log(4\pi) + 2\log D_{\rm L}(z) + \log F_0,
\end{equation}
in which clearly $a$ positively correlates with $\log H_0$. 

For uncalibrated correlations, the best-fit parameters can be found by maximizing the log-likelihood function  
\begin{equation}
\label{L2c}
\mathcal F_C = - \frac{1}{2} \sum_{i=1}^{\mathcal N_C} \! \left[\! \frac{ \left(\log \mathcal L_{0,i}\!-\!\log L_{0,i}\right)^2}{\sigma_C^2}\! + \ln(2 \pi \sigma_C^2 ) \!\right]\!, 
\end{equation}
where $\sigma_C^2 = \sigma_{\log F_{0,i}}^2 +  b^2 \sigma_{\log E_{\rm p,i}}^2 + \sigma_{\log T_i}^2 + \sigma_{\rm ex}^2$ is the global error of the correlation, and $\sigma_{\log F_0}$ and $\sigma_{\log T}$ are the errors of the logarithms of $F_0$ and $T$, respectively.
The results of the Monte Carlo Markov chain fit performed on the C182 data set are summarized in Table~\ref{tab1}.

\subsection{Dipole determination methodology}
\label{dip_proc}

To explore variations $\Delta a$ across the sky, we scan over right ascension $\alpha$ and declination $\delta$ considering a grid of $j$ points, each covering an angular patch of $12^\circ \times 12^\circ$. 
For each point on this grid we define the ortogonal vector 
\begin{equation}
\label{othov}
\vec{v}_j = \left(\cos\delta\,\cos\alpha\,,\cos\delta\,\sin\alpha\,,\sin\delta \right)\,.  
\end{equation}
Based on the sign of the inner product between $\vec{v}_j$ and the vector $\vec{v}_{\rm GRB,i}$ of each burst of A118 or C182 catalogs, for each point of the grid, we perform the steps below.
\begin{itemize}
\item[-] We split both A118 and C182 catalogs into two hemispheres, the ``northern" one (N) for $\vec{v}_j\cdot\vec{v}_{\rm GRB,i}>0$,  and the ``southern" one (S) for $\vec{v}_j\cdot\vec{v}_{\rm GRB,i}<0$. Clearly, flipping of $180^\circ$ duplicates the results; we keep both evaluations for consistency.
\item[-] The bursts in each hemisphere shall fulfill the same correlation of the whole sample (A118 or C182), namely, we fix $b$ and $\sigma_{\rm ex}$ to the values listed in Table~\ref{tab1}, whereas $a$ and $\Omega_m$ are let free, since $\Omega_m$ degenerates with $H_0$ and $a$ is a proxy for $H_0$. 
\item[-] In each hemisphere, we maximize $\mathcal F_A$ for the A118 catalog and $\mathcal F_C$ for the C182 sample, and find out the corresponding normalizations $a_{{\rm N},j}$ and $a_{{\rm S},j}$. 
\item[-] For both correlations, we compute the significance due to the variation of $a$ between hemispheres as
\begin{equation}
\label{sigma}
\sigma_j = \frac{a_{{\rm N},j} - a_{{\rm S},j}}{\sqrt{\sigma_{a_{{\rm N},j}}^2+\sigma_{a_{{\rm S},j}}^2}}.
\end{equation} 
\end{itemize}
Repeating the scan over all the points of the grid, one can obtain the Mollweide projection map of the significance $\sigma_j$ and compare it to the CMB dipole direction. 

Differently from Ref.~\cite{2022PhRvD.105j3510L}, we here utilize the full A118 catalog without (a) applying specific weights and (b) excluding GRBs around the CMB dipole direction. This choice preserves the integrity of the catalog and does not introduce any unwanted bias that may originate any artificial dipole. 
Moreover, we perform a comparison with the C182 catalog, which is derived from a different GRB correlation. Results are shown in Table~\ref{tab1}.

In the following, we also show that the above procedure is able to unravel the presence of a possible dipole if embedded in the data. 

\begin{table*}
\caption{Best-fit parameters and $1\sigma$ errors of $a$, $b$, $\sigma_{\rm ex}$ and $\Omega_m$ for both correlations, obtained for the following directions: whole sky, maximum $\Delta a$, and CMB dipole. For the last two directions, we indicated the equatorial coordinates $(\alpha,\delta)$ and the associated significance $\sigma$ only for the N hemisphere. The parameters $b$ and $\sigma_{\rm ex}$ exhibit errors only in the whole sky fits, when determined for the first time; in the other cases these parameters are fixed to the whole-sky values, as the correlation of the whole sample shall hold in each hemisphere.}
\centering
\setlength{\tabcolsep}{.85em}
\renewcommand{\arraystretch}{1.2}
\begin{tabular}{llllccccr}
\hline\hline
Correlation & Direction & $(\alpha,\delta)$ & Hemisphere & 
$a$ & $b$ & $\sigma_{\rm ex}$ & $\Omega_m$ &
$\sigma$
\\ 
\hline   
$E_{\rm p}-E_{\rm iso}$ & Whole sky & & $-$ & 
$50.07_{-0.39}^{+0.37}$  &
$1.11_{-0.13}^{+0.12}$   &
$0.41_{-0.04}^{+0.05}$   &
$0.61_{-0.32}^{+0.33}$   &
$-$\\
& $\Delta a_{\rm max}$ & $(144^\circ,-54^\circ)$ & N  &
$50.22^{+0.18}_{-0.18}$  &
$1.11$ & $0.41$ &
$0.31^{+0.22}_{-0.22}$  &
$0.79$\\
& & & S  &
$50.04^{+0.18}_{-0.18}$  &
$1.11$ & $0.41$ &
$0.53^{+0.26}_{-0.26}$  & \\
& CMB & $(168^{\circ}, -7^{\circ})$ & N &
$50.09^{+0.18}_{-0.18}$  & 
$1.11$ & $0.41$ &
$0.78^{+0.24}_{-0.24}$   &
$-0.03$\\
& & & S &
$50.10^{+0.20}_{-0.20}$  &
$1.11$ & $0.41$ &
$0.39^{+0.26}_{-0.26}$   & \\
\hline
$L_0-E_{\rm p}-T$ & Whole sky & & $-$ & 
$49.55_{-0.24}^{+0.29}$  &
$0.79_{-0.12}^{+0.09}$   &
$0.37_{-0.04}^{+0.04}$   &
$0.93_{-0.24}^{+0.07}$   &
$-$\\
& $\Delta a_{\rm max}$ & $(270^\circ,-6^\circ)$ & N  &
$49.64^{+0.13}_{-0.13}$  &
$0.79$ & $0.37$ &
$0.70^{+0.20}_{-0.20}$  &
$0.42$\\
& & & S  &
$49.65^{+0.14}_{-0.14}$  &
$0.79$ & $0.37$ &
$0.84^{+0.21}_{-0.21}$  & \\
& CMB & $(168^{\circ}, -7^{\circ})$ & N &
$49.65^{+0.13}_{-0.13}$  & 
$0.79$ & $0.37$ &
$0.72^{+0.18}_{-0.18}$   &
$0.13$\\
& & & S &
$49.50^{+0.14}_{-0.14}$  &
$0.79$ & $0.37$ &
$0.89^{+0.21}_{-0.21}$   & \\
\hline
\end{tabular}
\label{tab1}
\end{table*}

\subsection{Mock catalogs with artificial $H_0$ dipoles}
\label{appA}

To test significance and repeatability of the results in Table ~\ref{tab1}, we create mock $E_{\rm p}$--$E_{\rm iso}$ and $L_0$--$E_{\rm p}$--$T$ catalogs composed of $\mathcal N_A=118$ and $\mathcal N_C=182$ GRBs, respectively, both characterized by an artificial dipole. 
Precisely, we aim to prove that using  Eq.~\eqref{sigma}, indeed, the maximum (minimum) significance is found in the direction (anti-direction) of the dipole. 

\begin{figure*}[t]
\centering
\includegraphics[width=0.75\hsize,clip]{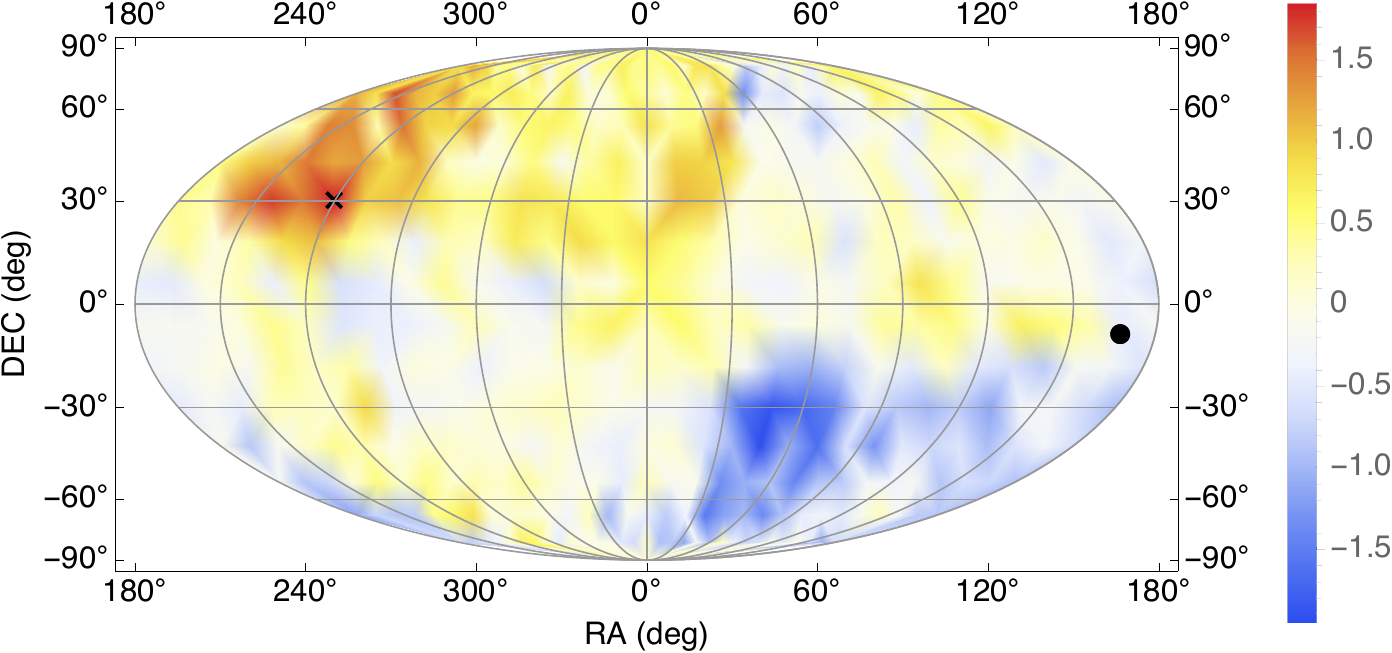}
\includegraphics[width=0.75\hsize,clip]{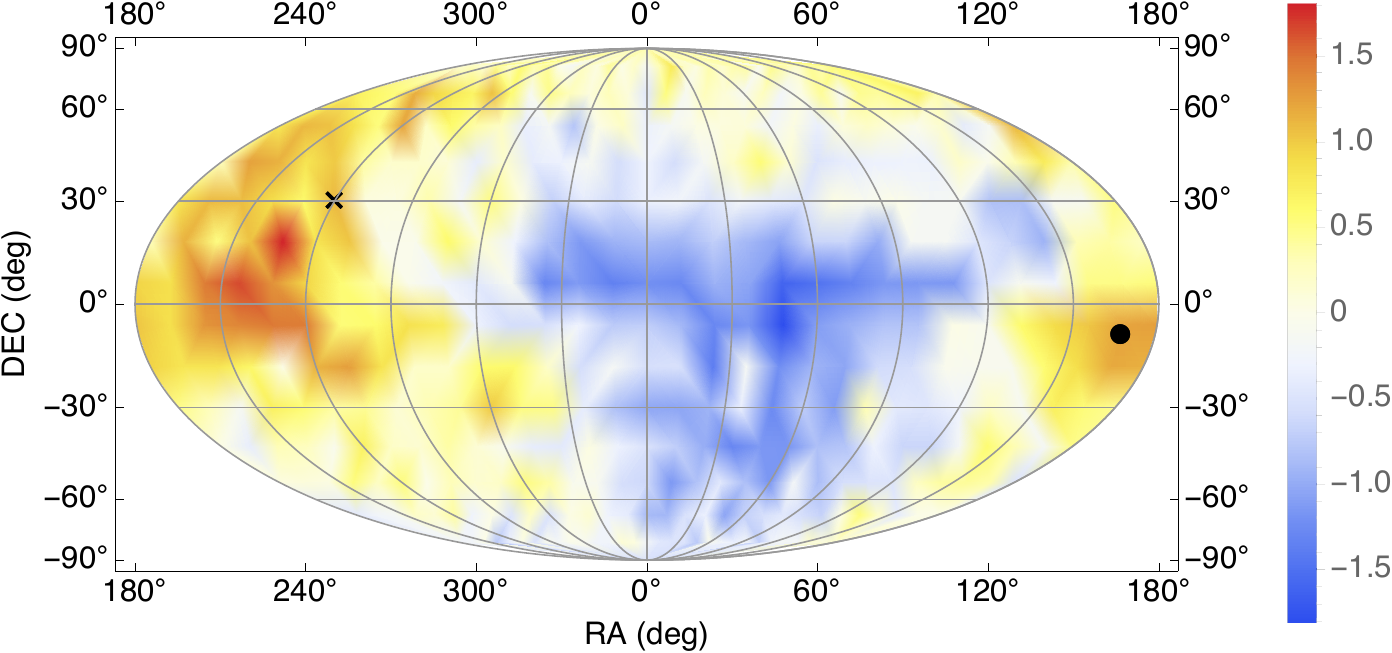}
\caption{Mollweide RA--DEC projection maps of the significance $\sigma$ (see right side color-coded bars) computed from Eq.~\eqref{sigma} for the mock A118 (top) and the C182 (bottom) catalogs. The artificial dipole direction $(\alpha_0,\delta_0) = (240^{\circ}, 30^{\circ})$ (black crosses) and the CMB dipole direction $(\alpha_\star,\delta_\star) = (168^{\circ}, -7^{\circ})$ (black dot) are also shown.}
\label{fig2}
\end{figure*}

The realizations of the mock catalogs of the $\mathcal N_A=118$ $E_{\rm p}$--$E_{\rm iso}$ GRBs and the $\mathcal N_C=182$ $L_0$--$E_{\rm p}$--$T$ bursts are based on the steps summarized below, some of which are taken from the procedure outlined in Ref.~\cite{Luongo:2024pcp}.
\begin{itemize}
\item[1.] We fix the direction of the artificial $H_0$ dipole at $(\alpha_0,\delta_0)=(240^\circ,30^\circ)$ and indicate the corresponding orthogonal vector defined by Eq.~\eqref{othov} with $\vec{v}_0$.
\item[2.] GRBs are isotropically distributed in the sky \cite{2022PhRvD.105j3510L}, thus, we produce $\mathcal N_A$ and $\mathcal N_C$ randomly distributed angular positions $(\alpha_k,\delta_k)$, for the $E_{p}-E_{iso}$ and the $L_0-E_{\rm p}-T$ correlations, respectively.
\item[3.] We compute the observed A118 isotropic energies $\log \mathcal E_{{\rm iso},i}$, through Eqs.~\eqref{eq2}--\eqref{Dlambda}, and the luminosities $\log \mathcal L_{0,i}$, through Eqs.~\eqref{Dlambda} and \eqref{l0}, using the $\Lambda$CDM paradigm best-fit values \citep{Planck2018}. This choice does not introduce any circularity in the procedure, since in the final catalog we generate $\log S_{{\rm b},k}$, for the $E_{p}-E_{iso}$ correlation, and $\log F_{0,k}$, for the Combo correlation, which are cosmology-independent.
\item[4.] To simplify the procedure, we MCMC fit the correlation $\log E_{{\rm p},i} = \bar m \log \mathcal E_{{\rm iso},i} + \bar q$ with dispersion $\bar \sigma_A$, built up from the A118 catalog, and derive the correlation parameters 
\begin{subequations}
\label{parsLCDM}
\begin{align}
&\bar m=\,0.537_{-0.039}^{+0.036}\,,\\
&\bar q=\,2.043^{+0.057}_{-0.055}\,,\\
&\bar \sigma_A =\,0.257_{-0.018}^{+0.018}\,.
\end{align}
\end{subequations} 

For the Combo correlation, we do not change its functional form, i.e. $\log L_{0,i}=\bar{a} + \bar{b}\log E_{{\rm p},i} - \log T_i$ (with dispersion $\bar{\sigma}_C$), thus we derive 
\begin{subequations}
\label{parsLCDMC}
\begin{align}
&\bar a=\,49.651_{-0.197}^{+0.200}\,,\\
&\bar b=\,0.841^{+0.075}_{-0.077}\,,\\
&\bar \sigma_C =\,0.377_{-0.022}^{+0.026}\,.
\end{align}
\end{subequations} 
\item[5.] We first fit the observed redshift distributions $\log z_i$ of both correlations with normal distributions.
From the A118 catalog we derive a normal distribution with mean $\mu_z=0.359$ and variance $\sigma_z=0.214$, and generate $\mathcal N_A$ mock $\log z_k$; from the C182 catalog we derive $\mu_z=0.267$ and variance $\sigma_z=0.281$, and generate $\mathcal N_C$ mock $\log z_k$.
\item[6.] Then, we focus on the independent variables of both correlations. 
For the Amati correlation, we fit all $\log \mathcal E_{{\rm iso},i}$ from A118 with a normal distribution. We deduce for this distribution a  mean $\mu_{\mathcal E}=53.300$ and a variance $\sigma_{\mathcal E}=0.718$, and use them to generate $\mathcal N_A$ mock values of $\log E_{{\rm iso},k}$. Reversing Eq.~\eqref{eq2}, it is possible to evaluate the corresponding mock logarithms of the bolometric fluences $\log S_{{\rm b},k}$.

For the Combo correlation, we fit all the quantities $\log X_i = \log E_{{\rm p},i}-\bar{b}^{-1}\log T_i$ from the C182 catalog with a normal distribution and deduce a mean $\mu_X=-1.362$ and a variance $\sigma_X=1.193$. Using them, we generate $\mathcal N_C$ mock values of $\log X_{k}$.
\item[7.] We attach the above generated angular positions $(\alpha_k,\delta_k)$ -- that define the corresponding orthogonal vectors $\vec{v}_{\rm GRB,k}$ -- to the pairs ($\log z_k$, $\log E_{{\rm iso},k}$) for the Amati correlation and ($\log z_k$, $\log X_k$) for the Combo correlation and, based on the sign of the inner product $\vec{v}_0\cdot\vec{v}_{\rm GRB,k}$, we split both the $\mathcal N_A$ and $\mathcal N_C$ mock catalogs into two subsamples corresponding to the hemispheres defined by the direction $\vec{v}_0$.
\item[8.] We generate artificial dipoles in both correlations.
For the $E_{\rm p}-E_{\rm iso}$ correlation we (a) increase the normalization to $\bar q+\delta\bar q$ in the N hemisphere, and (b) decrease it to $\bar q-\delta\bar q$ in the S hemisphere. Then, using Eqs.~\eqref{parsLCDM}, we generate the mock $\log E_{{\rm p},k}$ from normal distributions (a) with mean $\mu_{\rm p}^{\rm N}= \bar m \log E_{{\rm iso},k} + \bar q + \delta \bar q$ and variance $\bar \sigma_A$ for the N hemisphere, and (b) with mean $\mu_{\rm p}^{\rm S}= \bar m \log E_{{\rm iso},k} + \bar q - \delta \bar q$ and variance $\bar \sigma_A$ for the S hemisphere.

For the $L_0-E_{\rm p}-T$ correlation, the increased normalization in the N hemisphere is $\bar a+\delta\bar a$, whereas the decreased one in the S hemisphere is $\bar a-\delta\bar a$. 
Thus, using Eqs.~\eqref{parsLCDMC}, we generate the mock $\log L_{0,k}$ from normal distributions (a) with mean $\mu_{\rm L}^{\rm N}= \bar a + \delta \bar a + \bar b \log X_k$ and variance $\bar \sigma_C$ for the N hemisphere, and (b) with mean $\mu_{\rm L}^{\rm S}= \bar a - \delta \bar a + \bar b \log X_k$ and variance $\bar \sigma_C$ for the S hemisphere. Reversing Eq.~\eqref{l0}, we evaluate the corresponding mock of the X-ray fluxes $\log F_{0,k}$.

We use $\delta\bar q=1.5\bar\sigma_A$ and $\delta\bar a=1.5\bar\sigma_C$, so we expect dipoles in the direction $\vec{v}_0$ with a significances approximately $+1.5\sigma_A$ and $+1.5\sigma_C$, respectively.  
\item[9.] Finally, we generate the errors of the mock catalogs. 
For the mock A118 catalog, we determine the errors using the following weighted expressions
\begin{subequations}
\begin{align}
\label{errorEisosim}
\sigma_{\log E_{{\rm iso},k}} &= \langle \sigma_{\log E_{{\rm iso},i}}\rangle \frac{\log E_{{\rm iso},k}}{\langle \log E_{{\rm iso},i}\rangle},\\
\label{errorEpsim}
\sigma_{\log E_{{\rm p},k}} &= \langle \sigma_{\log E_{{\rm p},i}}\rangle \frac{\log E_{{\rm p},k}}{\langle \log E_{{\rm p},i}\rangle},
\end{align}
\end{subequations}
where the averaged values $\langle \sigma_{\log E_{{\rm p},i}}\rangle$, $\langle \log E_{{\rm p},i}\rangle$, $\langle \sigma_{\log E_{{\rm iso},i}}\rangle$ and $\langle \log E_{{\rm iso},i}\rangle$ are evaluated from the original A118 catalog. The errors in Eq.~\eqref{errorEisosim} are equivalent to the errors on $\log S_{{\rm b},k}$.

For the mock C182 catalog, the errors are obtained from the following weighted expressions
\begin{subequations}
\begin{align}
\label{errorXsim}
\sigma_{\log X_k} &= \langle \sigma_{\log X_i}\rangle \frac{\log X_k}{\langle \log X_i\rangle},\\
\label{errorL0sim}
\sigma_{\log L_{0,k}} &= \langle \sigma_{\log L_{0,i}}\rangle \frac{\log L_{0,k}}{\langle \log L_{0,i}\rangle},
\end{align}
\end{subequations}
where the averaged values $\langle \sigma_{\log X_i}\rangle$, $\langle \log X_i\rangle$, $\langle \sigma_{\log L_{0,i}}\rangle$ and $\langle \log L_{0,i}\rangle$ are evaluated from the original C182 catalog. The errors in Eq.~\eqref{errorL0sim} are equivalent to the errors on $\log F_{0,k}$.
\item[10.] In the last step, we apply the procedure outlined in Sec~.\ref{dip_proc} to the so-generated catalogs, i.e., $(\alpha_k,\delta_k,\log E_{{\rm p},k},E_{{\rm iso},k})$ for the mock A118 and $(\alpha_k,\delta_k,\log X_k,L_{0,k})$ for the mock C182.
Thus, we compute the significances $\sigma_k$ from Eq.~\eqref{sigma} over all the sky and for both correlations. The obtained Mollweide projection maps are shown in Fig.~\ref{fig2} and the results are summarized in Table~\ref{tab2}. 
\end{itemize}

\section{Physical discussions and conclusions}
\label{concl}

Following the above-described  procedure, we obtained the Mollweide projection maps shown in Fig.~\ref{fig1}. 
Table~\ref{tab1} shows the best-fit parameters focusing on the directions of maximum $\Delta a$, for both A118 and C182 catalogs. Comparison with the CMB dipole direction are also displayed.

From the above results we can conclude that:
\begin{itemize}
\item[-] there is no $H_0$ dipole in the CMB direction, where the significance drops to $-0.03\sigma$ for the Amati correlation and to $0.13$ for the Combo correlation;
\item[-] low-significance dipoles emerge from $\Delta a_{\rm max}$ directions, namely $(\alpha_{\rm m},\delta_{\rm m})=(144^\circ,-54^\circ)$ at $0.79\sigma$ from the A118 catalog, and $(\alpha_{\rm m},\delta_{\rm m})=(270^\circ,-6^\circ)$ at $0.42\sigma$ for the C182 catalog;
\item[-] the directions $(\alpha_{\rm m},\delta_{\rm m})$ for both correlations are inconsistent with $(\alpha_\star,\delta_\star)$, being the offsets $|\alpha_{\rm m}-\alpha_\star|$ and $|\delta_{\rm m}-\delta_\star|$ way larger than the numerical grid angular step of $12^\circ$.
\end{itemize}

\begin{table*}
\caption{Best-fit parameters and $1\sigma$ errors of $a$, $b$, $\sigma_{\rm ex}$ and $\Omega_m$ for the mock catalogs of both correlations, obtained for the following directions: maximum $\Delta a$, and CMB dipole. We indicated the equatorial coordinates $(\alpha,\delta)$ and the associated significance $\sigma$ only for the N hemisphere. The parameters $b$ and $\sigma_{\rm ex}$ do not exhibit errors as they are fixed to the whole-sky values from Table~\ref{tab1}, as the correlation of the whole sample shall hold in each hemisphere.}
\centering
\setlength{\tabcolsep}{1.1em}
\renewcommand{\arraystretch}{1.2}
\begin{tabular}{llllccccr}
\hline\hline
Correlation & Direction & $(\alpha,\delta)$ & Hemisphere & 
$a$ & $b$ & $\sigma_{\rm ex}$ & $\Omega_m$ &
$\sigma$
\\ 
\hline   
$E_{\rm p}-E_{\rm iso}$ & $\Delta a_{\rm max}$ & $(252^\circ,30^\circ)$ & N  &
$50.82^{+0.24}_{-0.24}$  &
$1.11$ & $0.41$ &
$0.04^{+0.23}_{-0.04}$  &
$1.85$\\
& & & S  &
$50.29^{+0.19}_{-0.19}$  &
$1.11$ & $0.41$ &
$0.70^{+0.25}_{-0.25}$  & \\
& CMB & $(168^{\circ}, -7^{\circ})$ & N &
$50.31^{+0.21}_{-0.21}$  & 
$1.11$ & $0.41$ &
$0.70^{+0.28}_{-0.28}$   &
$-0.46$\\
& & & S &
$50.36^{+0.22}_{-0.22}$  &
$1.11$ & $0.41$ &
$0.28^{+0.28}_{-0.28}$   & \\
\hline   
$L_0-E_{\rm p}-T$ & $\Delta a_{\rm max}$ & $(240^\circ,18^\circ)$ & N  &
$49.86^{+0.15}_{-0.15}$  &
$0.79$ & $0.37$ &
$0.04^{+0.12}_{-0.04}$  &
$1.81$\\
& & & S  &
$49.36^{+0.12}_{-0.12}$  &
$0.79$ & $0.37$ &
$0.65^{+0.20}_{-0.20}$  & \\
& CMB & $(168^{\circ}, -7^{\circ})$ & N &
$49.81^{+0.17}_{-0.17}$  & 
$0.79$ & $0.37$ &
$0.05^{+0.15}_{-0.15}$   &
$1.24$\\
& & & S &
$49.37^{+0.15}_{-0.15}$  &
$0.79$ & $0.37$ &
$0.51^{+0.24}_{-0.24}$   & \\
\hline
\end{tabular}
\label{tab2}
\end{table*}

To test the significance and validate the above results, in Sec.~\ref{appA} we generated mock A118 and C182 catalogs composed of $\mathcal N_A$ and $\mathcal N_C$ sources, respectively, in which an artificial dipole is injected in the direction $(\alpha_0,\delta_0)=(240^\circ,30^\circ)$.
The goal of this test is to prove that, if an intrinsic dipole exists in both catalogs, then the significance $\sigma_k$ computed from  Eq.~\eqref{sigma} shall be maximum and positive in the direction of the injected dipole.

As expected, dipole emerged in the directions $(\alpha_d,\delta_d)=(252^\circ,30^\circ)$ for the mock A118 catalog and $(\alpha_d,\delta_d)=(240^\circ,18^\circ)$ for the mock C182 catalog.
Both inferred directions are compatible with the injection direction $(\alpha_0,\delta_0)=(240^\circ,30^\circ)$, within angular offsets $|\alpha_d-\alpha_0|\leq12^\circ$ and $|\delta_d-\delta_0|\leq12^\circ$ due to the numerical grid angular steps.
In both catalogs, the significance of dipole detections, respectively $1.85\sigma$ and $1.81\sigma$, are compatible with the expectations. In this way, we have developed and tested a pipeline that can be easily applied to future larger GRBs datasets.

The results of Fig.~\ref{fig2} and Table~\ref{tab2} not only validate the mock GRB data, but also the method of Sec~.\ref{sec:GRB_dip}. 
Phrasing it differently, \emph{if a significant dipole is embedded in the catalog, it manifests itself -- within a numerical offset due to the grid step -- in the direction of the maximum positive significance $\sigma_k$ computed from Eq.~\eqref{sigma}}.

We conclude this work with what summarized below.
\begin{itemize}
\item[-] There is no evidence for a significant $H_0$ dipole in the two considered GRB catalogs. 
\item[-] The absence of a significant dipole in $H_0$, in the CMB dipole direction or in any other directions, from high-redshift sources like GRBs confirms the cosmological principle and, indirectly, excludes the presence of intrinsic dipoles.
\item[-] Both the A118 and the C182 catalogs (and, in general, all GRB catalogs) are much smaller than those of SNe Ia, quasars, etc... and are affected by large uncertainties. 
This effect, in principle, may hinder weak $H_0$ dipole detections.
\item[-] Both A118 and C182 catalogs do not have local bursts thus, unlike SNe Ia and quasars, GRBs are not sensitive to local peculiar velocities.
\item[-] The sources in A118 are isotropically distributed in the sky, whereas SNe Ia and quasars are not \cite{2022PhRvD.105j3510L}. It would be interesting to test if the dipole still exists simulating isotropic catalogs of SNe Ia and quasars.
\end{itemize}

Future efforts are needful to clarify with intermediate data points the presence or absence of the cosmic dipole. The use of GRBs is also influenced by the employed data set. Grid scanning the sky distribution of additional catalogs appears therefore necessary to go beyond our present approach. Additional studies are also quite important in view of clarifying the possible existence of the cosmic dipole by adopting scenarios that can predict the existence of dipoles. Our outcomes did not reproduce the more pronounced dipole signal reported in previous analyses, see e.g.~Ref.~\cite{2022PhRvD.105j3510L}, highlighting the need for a more thorough assessment of how different data sets may affect the dipole detectability. Last but not least, our analysis was worked out investigating the CMB dipole direction, focusing on $H_0$ and GRBs appear isotropic placed in the sky. Hence, additional studies are needful to understand and to simulate how much the degree of anisotropy would influence the net dipole, in order to infer the fundamental properties behind its possible presence.

\begin{acknowledgments}
OL acknowledges support by the  Fondazione ICSC, Spoke 3 Astrophysics and Cosmos Observations National Recovery and Resilience Plan (PNRR) Project ID CN00000013 ``Italian Research Center on  High-Performance Computing, Big Data and Quantum Computing" funded by the Italian Ministry of University and Research (MUR) - Mission 4, Component 2, Investment 1.4: Strengthening research structures and creation of "national R\&D champions" (M4C2-19)" - Next Generation EU (NGEU).
MM acknowledges support by the European Union - NGEU, Mission 4, Component 2, under the MUR - Strengthening research structures and creation of "national R\&D champions" on some Key Enabling Technologies - grant CN00000033 - NBFC - CUPJ13C23000490006. FS acknowledges financial support from the Swiss National Science Foundation
\end{acknowledgments}

%


\end{document}